\documentclass[conference]{IEEEtran}

\ifCLASSINFOpdf

\else

\fi

\usepackage{url}
\usepackage{graphicx}
\usepackage{epsfig}
\usepackage{ifmtarg}
\usepackage[font=small,labelsep=none]{caption}
\usepackage[font=footnotesize]{caption}
\usepackage{footnote}
\usepackage{cite}
\usepackage{amssymb}
\usepackage{amsthm}
\usepackage[fleqn]{amsmath}
\usepackage{amsmath}
\usepackage{amsfonts}
\usepackage{epstopdf}
\usepackage{epsfig}
\usepackage{algorithmic}
\usepackage[linesnumbered,ruled]{algorithm2e}
\usepackage[font=small,labelsep=none]{caption}
\usepackage[font=footnotesize]{caption}

\begin{document}
\title{Post-disaster 4G/5G Network Rehabilitation using Drones: Solving Battery and Backhaul Issues}
\author{
\IEEEauthorblockN{\large Mohamed Y. Selim and Ahmed E. Kamal} \\
\vspace{-.15 in}
\small
\IEEEauthorblockA{Iowa State University, Iowa State, USA, Email: \{myoussef, kamal\}@iastate.edu}\\ \vspace{-.15 in}
\vspace{-.15 in}
\vspace{-.15 in}
\vspace{-.15 in}
\vspace{-.15 in}
\vspace{-.15 in}
}
\maketitle
\begin{abstract}
Drone-based communications is a novel and attractive area of research in cellular networks. It provides several degrees of freedom in time (available on demand), space (mobile) and it can be used for multiple purposes (self-healing, offloading, coverage extension or disaster recovery). This is why the wide deployment of drone-based communications has the potential to be integrated in the 5G standard. In this paper, we utilize a grid of drones to provide cellular coverage to disaster-struck regions where the terrestrial infrastructure is totally damaged due to earthquake, flood, etc. We propose solutions for the most challenging issues facing drone networks which are limited battery energy and limited backhauling. Our proposed solution based mainly on using three types of drones; tethered backhaul drone (provides high capacity backhauling), untethered powering drone (provides on the fly battery charging) and untethered communication drone (provides cellular connectivity). Hence, an optimization problem is formulated to minimize the energy consumption of drones in addition to determining the placement of these drones and guaranteeing a minimum rate for the users. The simulation results show that we can provide unlimited cellular service to the disaster-affected region under certain conditions with a guaranteed minimum rate for each user.
\end{abstract}
\IEEEpeerreviewmaketitle
\section{Introduction}

Natural disasters always cause massive unpredictable loss to life and property. Various types of natural
disasters, such as geophysical, hydrological, climatological
and meteorological, among others, have caused losses of many
lives in addition to increase in material losses.
This is why the occurrence of natural disasters is a terrible problem irritating the whole world including both developed and developing countries \cite{Erdelj}.

Currently, efforts are being made in three directions: 1) pre-disaster preparedness 2) disaster assessment 3) post-disaster response and recovery. The first two directions mainly depend on the recognition and forecast monitoring. The post-disaster stage mainly focuses on the rescue operation and facilitates the first responders' mission. In the USA, the Drone Integration Pilot Program was launched in November 2017 under presidential memorandum from the White House \cite{memorandum} to maximize the benefits of Unmanned Aerial Vehicles (UAVs) technologies for mitigating risks to public safety and security. This memorandum was issued after the successful mission of drones during the last two disasters: hurricane Irma in Florida and the wildfires in California. In Europe, ABSOLUTE project is aiming to use flying drones to enhance the ground network, especially for public safety and emergency situations \cite{ABSOLUTE}.

Drone-based communications is considered as a strong candidate to be used regularly in 5G. Moreover, 3GPP is planning to support non-terrestrial networks, i.e., drones/UAVs, in the second phase of the 5G new radio standard which is expected to appear in the 3GPP Rel-16 by mid-2019. 


There are two major ways to practically implement Drone BSs (DBSs); tethered and untethered DBSs. A tethered DBS means that a drone is connected by a cable that provides power and/or backhauling. Although it may sound uncanny for a drone to be tethered by a cable, this has many advantages such as a stable power source and hence unlimited flying time and ultra-high speed backhaul. All these advantages have encouraged well-known companies to test tethered DBSs, such as Facebook's ``tether-Tenna", AT\&T's ``Flying Cell-On Wings (COWS)", and EE's, UK's largest cellular operator, ``Air Masts" \cite{Qingqing}. Such a tethering feature also limits the operations of DBSs to taking off, hovering and landing only which in some cases is useful.

On the other hand, untethered DBSs rely on the onboard battery for powering up the platform. Although untethered DBSs have limited flying time, they have fully controllable mobility in 3D space. Also, untethered DBS can adjust its placement based on users distribution \cite{Fotouhi}.

In emergency zones, where the disaster causes total loss to the cellular infrastructure, the network has to be rapidly rehabilitated to facilitate and support the rescue operations of the first responders. We propose to use a grid of DBSs to cover the affected area to provide an alternate connectivity solution. By using the mentioned grid of DBSs, the main technical challenges to face are the difficulty to charge and backhaul these DBSs. Our proposed solution for the limited DBS battery issue is to use another drone to charge the DBSs on the fly. This special drone, we call it Powering Drone (PD), has on its platform a large capacity battery which is used to charge the DBSs on the fly. For solving the backhaul issue, we propose to use a tethered Backhaul Drone (tBD) which is powered and backhauled via cabling. In addition to solving these challenges, we introduce an optimization problem to minimize the energy consumption of the DBSs' network.

\addtolength{\topmargin}{0.1in}

\subsection{Literature Review}
Coexistence of drone grid with a totally inactive cellular network in a post-disaster situation has not been sufficiently investigated especially the unexplored issues: battery recharging and backhauling.

The authors in \cite{Naqvi} used UAVs in disaster-resilience where they present a disaster-struck scenario where they presented the trade-off between the altitude, beamwidth angles and the coverage area of the UAVs. However, the authors in \cite{Graven} are using drones to capture a full up-to-date 3D terrain elevation model of the disaster area. They also use drones to place sensors in that area to create an efficient wireless sensor network to aid first responders.

The authors in \cite{Selimm} present a novel framework to mitigate the effect of the failure of any BS in 5G networks using both DBSs and ground BSs. They showed that their proposed hybrid approach outperforms the conventional BS failure approach.




\section{System Architecture}

We consider a geographical area that experienced a natural disaster where 100\% of its terrestrial cellular network is out of service. A grid of drones is used to provide cellular coverage to the affected area where drones are connected to each other using hybrid FSO/RF links and one of the DBSs acquires the backhaul connection from a post-disaster tBD installed hundred of meters from the disaster area.

\begin{figure}
            \centering
        \includegraphics[width=3.2in, height=2.2in]{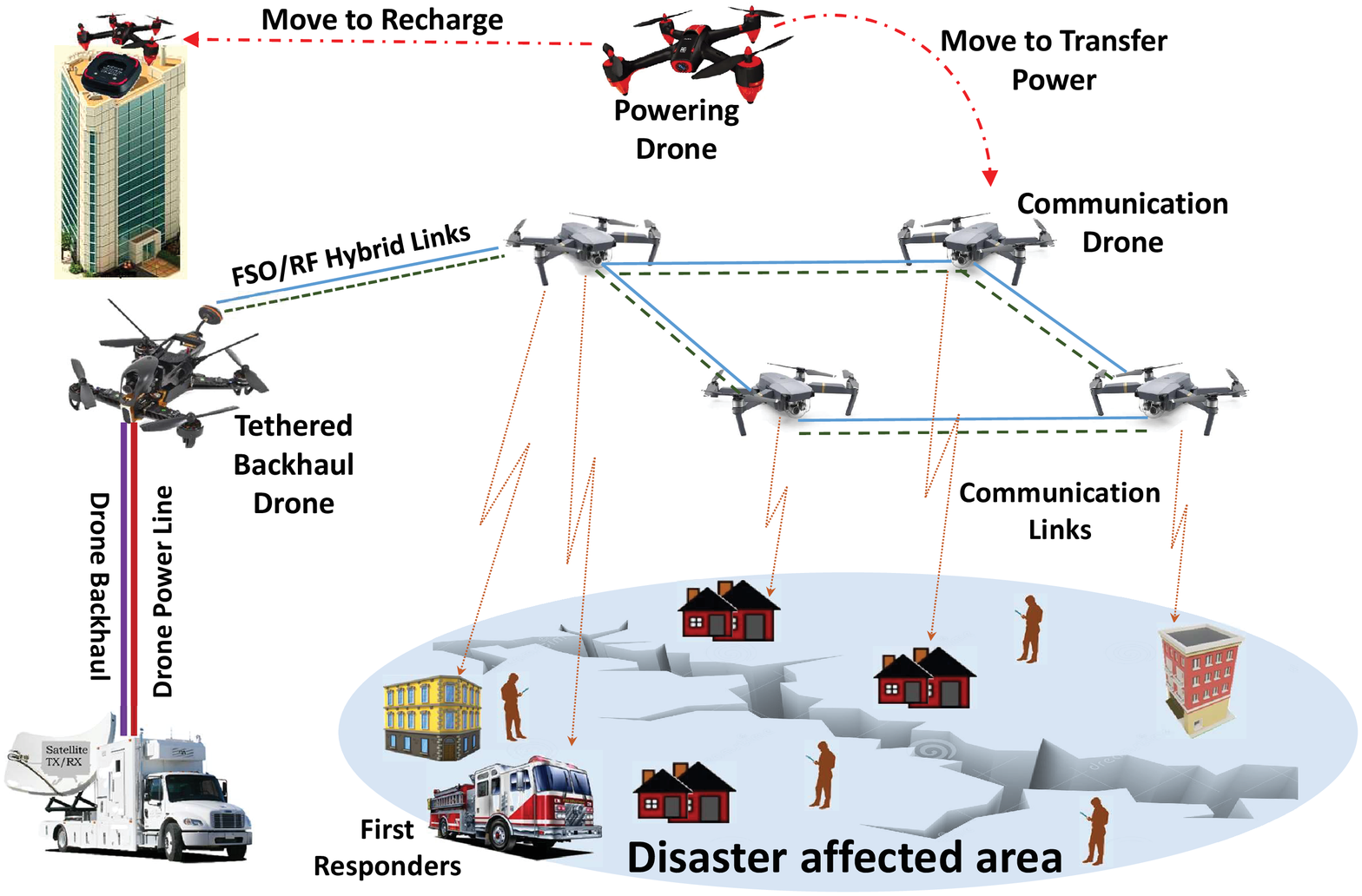}
        \caption{\,: System model during post-disaster rehabilitation.}
        \label{Sysmodel}
\vspace*{-0.5cm}
\end{figure}

Fig. \ref{Sysmodel}, shows the topology of the network during the post-disaster period. Upon the failure of the terrestrial BSs, the DBSs will fly to their initial positions to cover the whole footprint of the affected area. There are three types of drones: 1) tethered Backhaul Drone (tBD), 2) Powering Drone (PD), and 3) communication Drone BSs (cDBSs). The tBD provides the connectivity to the core network to the whole flying network via hybrid FSO/RF links. It also co-locates a central controller to manage and control this flying cellular network. The PD is mainly used to charge the cDBSs on the fly. This means that the cDBSs do not have to leave their locations to recharge their batteries. Finally, the cDBSs are used mainly to construct the flying cellular network to provide connectivity to first responders and users in the disaster area.

\subsection{cDBSs Backhauling}

During and post-disaster and especially when the whole cellular infrastructure is destroyed, the only available backhaul connection can be acquired from the satellite. However, in drone-based communications, it is impossible to equip the drone platform with satellite transceiver equipment. In our proposed solution and as it appears in Fig. \ref{Sysmodel}, we propose using a tBD which is powered and backhauled using a cable and this cable is connected to a special truck which is pre-equipped with satellite transceiver equipment to connect the drone cellular infrastructure to the core network.

The tBD provides backhauling to the DBSs grid via FSO/RF hybrid link. FSO has been constantly
claimed to be the alternative wireless technology of the
future that provides unlimited bandwidth. However, FSO is sensitive to atmospheric conditions apart
from precipitation. A solution to such a problem is to introduce a secondary wireless
channel which is less affected by such conditions like RF transmission. Commercial hybrid
FSO/RF systems have already made their presence
using a combination of millimeter Wave (mmW) and
laser-based FSO that allows more than 1 Gbps data
transmission over many kilometers of distance.

The reader is referred to \cite{Zedini} for understanding the performance analysis of the FSO/RF systems. In \cite{Alzenad}, the authors investigate the feasibility of a vertical backhaul framework where the UAVs transport the backhaul traffic between the access and core networks via FSO links.


\subsection{Drones Battery Charging}

The PD carries a large capacity battery, usually double the capacity of cDBS. This battery is used mainly to charge the untethered cDBSs whenever their batteries' charge is less than a certain threshold. The PD returns to its docking station to charge its battery while a replacement PD takes over the charging role so that the charging process will be available without any discontinuity. Finally, the untethered cDBS will have unlimited flying time due to the charge on the fly property provided by the PD. It is worth noticing that the charging process can be 1) wired charging and 2) inductive wireless charging. The wired charging has very high efficiency but it still needs special alignment technology between drones on the fly. The inductive wireless charging does not require physical contact since it can achieve an efficiency of 75\% given that the distance is within few inches. Hence, we consider the wired charging using advanced alignment techniques due to its high efficiency (almost 100\%).


\section{Problem Formulation}

In this architecture, a set $\mathcal{D}=\{1, 2, \dots , D\}$ of cDBSs are used to provide the needed coverage to the affected area. These cDBSs can dynamically move, when needed, to effectively mitigate the effect of the cellular infrastructure failure. The set $\mathcal{U}=\{1,2,\dots,U\}$ denotes the set of active UEs within the affected area and they are at known locations where the horizontal coordinate of each UE $u$ is fixed at $\mathbf{g}_u=[x_u,y_u]^T$, $u \in U$. All DBSs are assumed to navigate at a fixed altitude $h$ and the horizontal coordinate of DBS, $d$, at discrete time block $n$ where $n=1,...,N$ is denoted by $\mathbf{J}_d^n=[x_d^n,y_d^n]^\text{T}$ where $N$ is a total discrete period where $n$ are time blocks of equal duration $T_n$ and the total time is given by $T$. 

Assume that the DBS-UE communication channels are dominated by Line-of-Site (LoS) links. Though simplified, the LoS model offers a good approximation for practical Drone-UE channels and enables us to investigate the main objective of the optimization problem presented later. Given that $\mathbf{J}_d^n$ and $\mathbf{g}_u$ are the coordinates of DBS, $d$, and UE $u$ in the horizontal plane, respectively, then the distance from DBS, $d$, to UE $u$ during time block $n$ is given as $\delta_{u,d}^n=\sqrt{h^2+||\mathbf{J}_d^n-\mathbf{g}_u||^2}$.

\subsection{cDBS Channel and Achievable Rate Models}

The DBS-UE channel power gain mainly follows the free space path loss model which is given as follows:

\small
\begin{equation}\label{pathlssd}
  \Gamma_{u,d}^n=\rho_{o} (\delta_0/\delta_{u,d}^n)^{2}=\frac{\rho_{o}}{h^2+||\mathbf{J}_d^n-\mathbf{g}_u||^2}
\end{equation}
\normalsize

where $\rho_{o}$ is a unitless constant that depends on the antenna characteristics and frequency, and is measured at the reference distance $\delta_0=\text{1~m}$ and $\delta_{u,d}^n)$ is the square of the Euclidean distance between cDBS $d$ and user $u$.

Let $\mathcal{M}=\{1,2,\dots,M\}$ be the set of sub-channels that each DBS can use during the rehabilitation process. These sub-channels will be further divided and allocated to the UEs associated with each DBS. Each DBS, $d$, transmits to each UE, $u$, with a per sub-channel transmit power $p_{u,d,m}^n$. If sub-channel $m$ is not assigned to DBS, $d$, then $p_{u,d,m}^n$ will equal to zero. Hence, the SINR between DBS, $d$, and UE $u$ per sub-channel $m$ during time block $n$ can be expressed as:

\small
\begin{align}\label{SINRd}
  \gamma_{u,d,m}^n= & \frac{p_{u,d,m}^n~\Gamma_{u,d}^n}{\sum\limits_{\substack{i\in\mathcal{U} \\ i\neq u}}\sum\limits_{\substack{j\in\mathcal{D}}} p_{i,j,m}^n \Gamma_{u,j}^n + \sigma^2}
\end{align}
\normalsize

where $\sigma^2$ is the power of the Additive White Gaussian Noise at the receiver. 


Accordingly, the achievable per sub-channel downlink rate from DBS, $d$, to UE, $u$, is given by:
\vspace{-.05 in}
\small
\begin{equation}
  R_{u,d,m}^n=\text{log}_{\text{2}}(1+\gamma_{u,d,m}^n)
\end{equation}
\normalsize



\subsection{Drone Battery Energy Consumption Model}

In our proposed solution we have two types of untethered drones: 1) PD and 2) cDBS. Both of them consume hovering and hardware powers. We denote that the speed of the DBS $d$ in time block $n$ denoted by $v_d^n$. The hovering and hardware drone energy levels, denoted by $E_\text{hov}$ and $E_{\text{har},d}^n$, can be expressed, respectively, as \cite{wcnc}:

\small
\begin{equation}\label{EharEhov1}
  E_{\text{har},d}^n=\bigg[\frac{P_\text{full}-P_\text{idle}}{v_\text{max}}v_d^n+P_\text{idle}\bigg](T_\text{move})
\end{equation}
\normalsize

\small
\begin{equation}\label{EharEhov2}
  E_{\text{hov}}=\sqrt{\frac{(m_\text{tot} g)^3}{2 \pi r_p^2 n_p \rho}}~(T-T_\text{move})
\end{equation}
\normalsize

where $m_\text{tot}$, $g$, and $\rho$ are the drone mass in ($\text{Kg}$), earth gravity in (m$/\text{s}^2$), and air density in $(\text{Kg}/\text{m}^3)$, respectively. $r_p$ and $n_p$ are the radius and the number of the drone's propellers, respectively. $v_\text{max}$ is the maximum speed of the drone. $P_\text{full}$ and $P_\text{idle}$ are the hardware power levels when the drone is moving at full speed and when the drone is in idle mode, respectively. $T_\text{move}$ is the time used by cDBS to move from one location to another.

Hence, the total energy consumed by cDBSs is given as:

\small
\begin{equation}\label{energy}
  E=\sum_{d}\sum_{u}\sum_{m}\sum_{n} p_{u,d,m}^n~T + \sum_{d}\sum_{n} \big[E_{\text{har},d}^n + E_{\text{hov}}\big]
\end{equation}
\normalsize

Given that the initial battery level of DBS, $d$, is $B_0$, hence, the battery level of DBS, $d$, at time block $n$ is given by:

\small
\begin{align}
B_d^n&=B_0-\sum_{i=1}^{n} \big[ E_{\text{har},d}^i + E_{\text{hov}} + \sum_{u}\sum_{m} p_{u,d,m}^i~T \big]\nonumber\\
 &+ \sum_{d}\sum_{i=1}^{n} \beta_d^i(B_\text{charge})
\end{align}
\normalsize

where $\beta_d^n$ is a decision variable indicating whether PD is going to charge DBS, $d$, during time block $n$ or not. $B_\text{charge}$ represents the amount of charge that DBS, $d$, will receive from PD during one time block.

The PD battery model is different since it is not used for communication. Hence, it is given by:

\small
\begin{equation}\label{PDBattery}
  B_\text{PD}^n=B_{00}-\sum_{i=1}^{n} \big[ E_{\text{har},d}^i + E_{\text{hov}} \big]-\sum_{d}\sum_{i=1}^{n} \beta_d^i(B_\text{charge})
\end{equation}
\normalsize

where $B_{00}$ is the initial battery charge of PD. The term $\sum_{d}\sum_{i=1}^{n} \beta_d^i(B_\text{charge})$ represents the consumed energy up to time block $n$ used to charge the cDBSs.

\section{The Optimization Problem}

We formulate an optimization problem aiming to minimize the network's energy consumption during n time blocks.

We assume that initially the battery of the PD or the DBS is fully charged. Defining the decision variables: $\psi_{u,d}^n$ as the user association between user $u$ and cDBS $d$ during time block $n$ and $\Phi_{u,d,m}^n$ as the resource $m$ allocation to user $u$ by cDBS $d$ during time block $n$. Hence, the optimization problem minimizing the total energy consumption of the untethered cDBSs is given as:

\small
\begin{subequations}
\begin{align}
&\hspace{-0.5cm} (\textbf{P1}):\underset{{\mathbf{v}},{\mathbf{J}}, {\Phi}, {\Psi} , {\textbf{p}}}{\text{minimize}} \sum_{d}\sum_{u}\sum_{m}\sum_{n} \psi_{u,d}^n~ \Phi_{u,d,m}^n ~ p_{u,d,m}^n~T  \nonumber\\
&\hspace{-0.5cm} \quad\quad\quad\quad\quad\quad\quad+\sum_{d}\sum_{n} \big[E_{\text{har},d}^n + E_{\text{hov}}\big]  \label{of}\\
&\hspace{-0.5cm} \text{subject to:}\nonumber\\
&\hspace{-0.5cm}  \sum_{d}\sum_{m} \psi_{u,d}^n \Phi_{u,d,m}^n R_{u,d,m}^n \geq R^\text{th}         , \quad\quad\quad \forall~u,n      \label{minRate}\\
&\hspace{-0.5cm}  \sum_{u}\sum_{d}\sum_{m} \psi_{u,d}^n \Phi_{u,d,m}^n R_{u,d,m}^n \leq R^\text{BH} , \quad \forall~n \label{backhaul}\\
&\hspace{-0.5cm}  \sum_{d} \psi_{u,d}^n=1                          ,~~\quad\quad\quad\quad\quad\quad\quad\quad\quad\quad\quad \forall~u,n    \label{AssociationC}\\
&\hspace{-0.5cm}  \sum_{d}\sum_{m} \Phi_{u,d,m}^n \geq 1                          ,~~~\quad\quad\quad\quad\quad\quad\quad\quad \forall~u,n   \label{resources}\\
&\hspace{-0.5cm}  \beta_d^n \geq \frac{B^\text{th}-B_d^n}{Q}                      ,~\quad\quad\quad\quad\quad\quad\quad\quad\quad\quad \forall~ d,n \label{beta1}\\
&\hspace{-0.5cm}  \beta_d^n \leq \frac{B^\text{th}}{B_d^n}        ,\quad\quad\quad\quad\quad\quad\quad\quad\quad\quad\quad\quad\quad \forall~ d,n   \label{beta2}\\
&\hspace{-0.5cm}  \mathbf{J}_d^{\text{min}} \leq \mathbf{J}_d^n \leq \mathbf{J}_d^{\text{max}}   ,~~\quad\quad\quad\quad\quad\quad\quad\quad\quad \forall~ d,n \label{coordinates}    \\
&\hspace{-0.5cm}  ||\mathbf{J}_d^n-\mathbf{J}_d^{n-1}||=v_d^n ~T_\text{move}        , ~\quad\quad\quad\quad\quad\quad \forall~ d,n \label{velocitydistance}  \\
&\hspace{-0.5cm}  0 \leq v_d^n \leq v^\text{max}                        ,  \quad\quad\quad\quad\quad\quad\quad\quad\quad\quad\quad \forall~ d,n  \label{velocity}\\
&\hspace{-0.5cm} \sum_{u} \sum_{m} p_{u,d,m}^n \leq P^{\text{max}}                  , ~\quad\quad\quad\quad\quad\quad\quad \forall~ d,n        \label{maxpower} \\
&\hspace{-0.5cm} p_{u,d,m}^n \geq 0                , ~\quad\quad\quad\quad\quad\quad\quad\quad\quad\quad\quad\quad \forall~ u,d,m  \label{minimum power}\\
&\hspace{-0.5cm}  \psi_{u,d}^n,~\Phi_{u,d,m}^n~ ,\beta_d^n \in \{0,1\}   \quad\quad\quad\quad\quad\quad \forall~ u,d,m,n  \label{decisionVariables}
\end{align}
\end{subequations}
\normalsize

Constraint (\ref{minRate}) represents the QoS constraint on the rate of each use, $u$, where $R^\text{th}$ is the threshold rate. The backhaul constraint is given by (\ref{backhaul}). Constraint (\ref{AssociationC}) is limiting the association of each user to one cDBS only during each time block where $\psi_{u,d}^n$ is the association between cDBS $d$ and user $u$ during time block $n$. Constraint (\ref{resources}) guarantees that each user is getting at least one resource block. Constraints (\ref{beta1}) and (\ref{beta2}) together are enforcing $\beta_d^n$ to equal to $1$ if the PD is going to charge cDBS, $d$, during time block, $n$ where $Q$ is a very large number. This enforcement occurs if $B_d^n\leq B^\text{th}$ where $B^\text{th}$ is a certain threshold. Constraint (\ref{coordinates}) is limiting all cDBSs to fly within the disaster region. However, constraints (\ref{velocitydistance})-(\ref{velocity}) control the velocity and displacement of cDBSs. Finally, constraints (\ref{maxpower}) and (\ref{minimum power}) provides the minimum and maximum power limits of each cDBS.

$\bf{P1}$ is not easy to solve due to the decision variables $\Phi_{u,d}^m$, $\psi_{u,d}$ and $\beta_d^n$ and the non-convexity appearing in the objective function (\ref{of}), constraint (\ref{minRate}) and (\ref{backhaul}) with respect to cDBS coordinates and downlink power, $p_{u,d}^m$. Therefore, problem (\ref{of}) is difficult to be solved optimally. To make $\bf{P1}$ more tracktable, we propose to add the following constraint to $\bf{P1}$:

\begin{equation}\label{powerphipsi}
  p_{u,d}^m \leq \psi_{u,d} \Phi_{u,d}^m P^{\text{max}}, \quad \forall~ u,d,m
\end{equation}

Constraint (\ref{powerphipsi}) is used mainly to force $p_{u,d,m}^n$ to equal to zero if $\Phi_{u,d,m}^n$ and/or $\psi_{u,d}^n$ equal to zero. Consequently, there is no need to multiply the term $\psi_{u,d}^n\Phi_{u,d,m}^n$ by $p_{u,d,m}^n$ as done in the objective function. The same concept applies to constraints (\ref{minRate}) and (\ref{backhaul}).

Constraint (\ref{powerphipsi}) is non-linear. It can be linearized without any approximation by replacing it by the following three constraints:

\small
\begin{subequations}
\begin{align}
&\hspace{-0.5cm} p_{u,d,m}^n \leq \psi_{u,d}^n ~P^{\text{max}},  \quad\quad\quad\quad\quad\quad\quad\quad\quad\quad \forall~ u,d,m,n \label{linear1}\\
&\hspace{-0.5cm} p_{u,d,m}^n \leq \Phi_{u,d,m}^n ~P^{\text{max}}, \quad\quad\quad\quad\quad\quad\quad\quad\quad \forall~ u,d,m,n \label{linear2}\\
&\hspace{-0.5cm} p_{u,d,m}^n \geq (\psi_{u,d}^n+\Phi_{u,d,m}^n-1)~P^{\text{max}}, ~\quad\quad\quad \forall~ u,d,m,n \label{linear3}
\end{align}
\end{subequations}
\normalsize

After adding the new constraints and eliminating the non-linearity from the objective function of $\bf{P1}$ and eliminating $\psi_{u,d}^n$ and $\Phi_{u,d,m}^n$ from constraints (\ref{minRate}) and (\ref{backhaul}) and expanding $R_{u,d,m}^n$, we introduce $\bf{P2}$ which is a modified, non approximated, version of $\bf{P1}$ which is given as follows:

\small
\begin{subequations}
\begin{align}
&\hspace{-0.7cm} (\textbf{P2}): \underset{{\mathbf{v}},{\mathbf{J}}, {\Phi}, {\Psi} , {\textbf{p}}}{\text{minimize}} \sum_{d}\sum_{u}\sum_{m}\sum_{n} p_{u,d,m}^n~T +\sum_{d}\sum_{n} \big[E_{\text{har},d}^n + E_{\text{hov}}\big]  \label{of1}\\
&\hspace{-0.5cm} \text{subject to:}\nonumber\\
&\hspace{-0.5cm} ~~~~~~~~~~~~~~~~~\text{Constraints (\ref{AssociationC})~-~(\ref{decisionVariables}), (\ref{linear1})-(\ref{linear3})} \nonumber\\
&\hspace{-0.5cm}  \sum_{d}\sum_{m} \text{log}_{\text{2}}\big(1+\frac{p_{u,d,m}^n~\Gamma_{u,d}^n}{\sum\limits_{\substack{i\in\mathcal{U} \\ i\neq u}}\sum\limits_{\substack{j\in\mathcal{D}}} p_{i,j,m}^n \Gamma_{u,j}^n + \sigma^2}\big) \geq R^\text{th}         , ~~\quad \forall~u,n      \label{minRate1}\\
&\hspace{-0.5cm}  \sum_{u}\sum_{d}\sum_{m} \text{log}_{\text{2}}\big(1+\frac{p_{u,d,m}^n~\Gamma_{u,d}^n}{\sum\limits_{\substack{i\in\mathcal{U} \\ i\neq u}}\sum\limits_{\substack{j\in\mathcal{D}}} p_{i,j,m}^n \Gamma_{u,j}^n + \sigma^2}\big) \leq R^\text{BH} ,  \forall~n \label{backhaul1}
\end{align}
\end{subequations}
\normalsize

$\bf{P2}$ is still not easy to solve due to the binary variables $\Phi_{u,d}^m$ and $\psi_{u,d}$ and the non-linearity in constraints (\ref{minRate1}) and (\ref{backhaul1}).

For simplicity and given that the tBD has high speed backhaul wired link, we will consider that the backhaul rate is always greater than the sum rate of all users.

This simplicity assumption is supported by the simulation results from \cite{Alzenad}. In addition, we claim that post-disaster users are not using high bandwidth application(s) during this hard situation. Hence, we can ignore constraint (\ref{backhaul1}).

\section{The Proposed Solution}

In general, $\bf{P2}$ has no standard method for solving it efficiently. In the following, we propose an efficient iterative algorithm for solving $\bf{P2}$. Specifically, for a given coordinate $\mathbf{J}_d$, we optimize the decision variables $\beta_d^n$, $\Phi_{u,d}^m$ and $\psi_{u,d}$ and the continuous variable $p_{u,d,m}^n$ based on the Successive Convex Approximation (SCA) technique \cite{SCAICC}. Then for a given decision variables and power, we find the cDBSs coordinates using the same technique. Finally, a joint iterative algorithm is proposed to solve $\bf{P2}$ efficiently.

\subsection{Solving for cDBS Power and Decision Variables}\label{DBS1}

For any given coordinates, $\mathbf{J}_d$, the cDBS downlink power and decision variables of $\bf{P2}$ can be optimized by solving the following problem:

\small
\begin{align}
&\hspace{-0.7cm} (\textbf{P3}): \underset{{\mathbf{v}},{\Phi}, {\Psi} , {\textbf{p}}}{\text{minimize}} \sum_{d}\sum_{u}\sum_{m}\sum_{n} p_{u,d,m}^n~T +\sum_{d}\sum_{n} \big[E_{\text{har},d}^n + E_{\text{hov}}\big]  \label{of2}\\
&\hspace{-0.5cm} \text{subject to:}\nonumber\\
&\hspace{-0.5cm} ~~~~~~~~~~\text{Constraints (\ref{AssociationC})~-~(\ref{decisionVariables}), (\ref{linear1})~-~(\ref{linear3}), (\ref{minRate1})} \nonumber
\end{align}
\normalsize

$\bf{P3}$ is a non-convex optimization problem due to constraint (\ref{minRate1}). Based on the mathematical manipulation presented in \cite{RuiZhangTraj}, this constraint can be rewritten as:

\small
\begin{align}
&\sum_{m} \Big[\underbrace{\text{log}_{\text{2}}\big(\sum\limits_{\substack{i\in\mathcal{U}}}\sum\limits_{\substack{j\in\mathcal{D}}} p_{i,j,m}^n \Gamma_{u,j}^n + \sigma^2\big)}_{\tilde{R}^1_{u,m,n}} \nonumber\\
&-\underbrace{\text{log}_{\text{2}}\big(\sum\limits_{\substack{i\in\mathcal{U} \\ i\neq u}}\sum\limits_{\substack{j\in\mathcal{D}}} p_{i,j,m}^n \Gamma_{u,j}^n + \sigma^2\big)}_{\tilde{R}^2_{u,m,n}}\Big] \geq R^\text{th}, \quad \forall~ u  \label{R1R2}
\end{align}
\normalsize

From constraint (\ref{R1R2}), it can be noticed that this is a difference of two concave functions0. This difference is not guaranteed to be neither concave nor convex. This motivates us to approximate $\tilde{R}^2_{u,m}$. To convert constraint (\ref{R1R2}) to a convex one, we apply the SCA technique to approximate $\tilde{R}^2_{u,m,n}$ by a linear function in each iteration. Let $p_{u,d,m}^n(r)$ is the given cDBS power in the r-th iteration. Since any concave function is globally upper-bounded by its first-order Taylor expansion at any point \cite{RuiZhangTraj}. Thus, the second term of Eq. (\ref{R1R2}), can be upper bounded as follows:

\small
\begin{align}
\tilde{R}^2_{u,m,n}=&\text{log}_{\text{2}}\big(\sum\limits_{\substack{i\in\mathcal{U} \\ i\neq u}}\sum\limits_{\substack{j\in\mathcal{D}}} p_{i,j,m}^n \Gamma_{u,j}^n + \sigma^2\big)\nonumber\\
\leq & \sum\limits_{\substack{i\in\mathcal{U} \\ i\neq u}}\sum\limits_{\substack{j\in\mathcal{D}}} \frac{\text{log}_{\text{e}}\Gamma_{u,j}^n~(p_{u,d,m}^n-p_{u,d,m}^n(r))}{\sum\limits_{\substack{i\in\mathcal{U} \\ i\neq u}}\sum\limits_{\substack{j\in\mathcal{D}}} p_{i,j,m}^n(r) \Gamma_{u,j}^n + \sigma^2} \nonumber\\
+&\text{log}_{\text{2}}\big(\sum\limits_{\substack{i\in\mathcal{U} \\ i\neq u}}\sum\limits_{\substack{j\in\mathcal{D}}} p_{i,j,m}^n(r) \Gamma_{u,j}^n + \sigma^2\big) \overset{\Delta}{=} \tilde{\tilde{R}}^2_{u,m,n}
\end{align}
\normalsize

Constraint (\ref{minRate1}) is now convex, hence, $\bf{P3}$ is now convex which can be solved efficiently.

\subsection{Solving for cDBS Coordinates}\label{DBS2}




Solving $\bf{P2}$ for cDBSs coordinates $\mathbf{J}_d^n$ and fixing all other variables will result in aproblem which is not easy to solve. Using SCA in this case is not optimally efficient since we have to linearize both logarithmic functions if we expanded (\ref{coordinates1}) in the same way of constraint (\ref{R1R2}). It is proved in \cite{RuiZhangTraj} that linearizing/convexifying this constraint is not easy in general. This motivates us to find the cDBSs' coordinates using the following heuristic approach.

Due to the non-convexity of the problem even with fixed decision variables and downlink power, we introduce an efficient algorithm to find the optimal cDBSs' coordinates, $\textbf{J}_d$.

The algorithm starts by dividing the desired area into equal sectors based on the number of the cDBSs and each cDBS is placed initially in the middle of the sector. Then we generate certain number of particles in each sector to identify promising candidates and to form initial populations. Then, it determines the objective function achieved by selected particles by solving \textbf{P3}. After that, it finds the particle that provides the highest solution for this iteration.
Then, we generate a subset number of particles around this highest solution and calculate the objective function to find the best particle. This procedure is repeated until convergence or reach maximum iteration.


Algorithm 1 is an iterative efficient algorithm used to solve Problem $\bf{P2}$. Line 1 initiate the iteration and termination conditions. Lines 2-4 used to replace PD if its battery level is below the threshold then lines 5-7 make sure that the PD is charging only 1 DBS at each time block. Lines 8-9 solve $\bf{P3}$ for fixed cDBSs' location. By fixing the coordinates of the cDBSs and solving $\bf{P3}$ using SCA, then lines 10-13 generate particles and compute the objective function at each candidate point. From line 15 to 17 the algorithm finetunes the best placement by searching nearby particles for the best candidate coordinate and this is repeated at each iteration to find $l^{\text{r,local}}_d$ which indicates the index of the best local particle that results in the highest objective function for iteration $r$.

\begin{algorithm}[h!]
\caption{Joint optimization algorithm}\label{joint}
\small
\KwIn{ Initial positions for UAVs $\mathbf{J}_d^n(0)$}
\KwOut{$\mathbf{J}_d^n(r+1)$, $\psi_{u,d}^n(r+1)$, $\Phi_{u,d,m}^n(r+1)$, $\beta_d^n(r+1)$, $p_{u,d,m}^n(r+1)$, $v_d^n(r+1)$}
\begin{algorithmic}[1]
\WHILE {{Not} converged or reach maximum iteration}
\IF{$B_\text{PD}^n \leq B_\text{PD}^{\text{th}}$}
\STATE Replace PD
\ENDIF
\IF{$\sum_{d}\beta_d^n \geq 2$}
\STATE Choose cDBS $d$ randomly to be charged
\ENDIF
\STATE Solve \bf{P3} for the given $\mathbf{J}_d^n(r)$
\STATE Denote results as $p_{u,d,m}^n(r+1)$ and $\Phi_{u,d,m}^n(r+1)$
\STATE Generate initial population ${\mathcal L}$ composed of $L$ particles
\FOR {$l=1 \cdots L$}
\STATE Compute corresponding objective function of \bf{P4} \\given $\psi_{u,d}^n(r+1)$, $\Phi_{u,d,m}^n(r+1)$, $\beta_d^n(r+1)$, $p_{u,d,m}^n(r+1)$, $v_d^n(r+1)$
\ENDFOR
\STATE Find $(l^{\text{r,local}}_d)=\underset{l,d}{\arg\mathrm{min}}~ \sum_{d}\sum_{u}\sum_{m}\sum_{n} p_{u,d,m}^n~T +\sum_{d}\sum_{n} \big[E_{\text{har},d}^n + E_{\text{hov}}\big]$
\STATE Generate a subset of particles around $l^{\text{r,local}}_d$
\STATE Use shrink-and-realign sample spaces process to find \\the best solution i.e., $l^{\text{r,sub-optimal}}_d$
\STATE $l^{\text{r,local}}_d=l^{\text{r,sub-optimal}}_d$, $\forall d$ \text{and} $\mathbf{J}_d^n(r+1)=l^{\text{r,sub-optimal}}_d$
\STATE Update r=r+1.
\ENDWHILE
\end{algorithmic}
\normalsize
\end{algorithm}

\section{Numerical Results}

In this section, numerical results are provided to investigate the benefits of using cDBSs in mitigating disaster effects. The simulation area is 800x800 $\text{m}^\text{2}$ where the users are distributed randomly over this area given that all terrestrial ground BSs are inactive. Under the post-disaster scenario, we initialized 4 standby cDBSs to be used in the mitigation process. We use two PD where one is active and the other is standby in case its battery is depleted. Simulation was carried out using General Algebraic Modeling System (GAMS) https://www.gams.com/". GAMS is a high-level modeling system for mathematical programming and optimization. It is designed for modeling and solving linear, nonlinear, and mixed-integer optimization problems. GAMS is tailored for complex, large scale modeling problems, and allows to build large maintainable models. The parameters used in the simulation are presented in Table II. Also, the parameters of $E_\text{hov}$ and $E_\text{har}$ can be found in \cite{wcnc} given that $m_\text{tot}$ for PD is double that of cDBS.

The battery specifications of cDBS and PD are taken from a real market specifications. For cDBS it has 3cell battery with 11.1 volts, 5000 mAh and 55.5 Wh. The PD has a double battery capacity specifications where it has 6 cells with 22.2 volts, 10,000 mAh and 222 Wh.


\begin{table}[t]
\centering
\vspace*{0.15in}
\caption{\label{tab2} System parameters}
\addtolength{\tabcolsep}{-2pt}\begin{tabular}{|l|c||l|c||l|c|}
\hline
\textbf{Parameter}           & \textbf{Value}  & \textbf{Parameter}    & \textbf{Value} & \textbf{Parameter}          & \textbf{Value}\\ \hline \hline
$P_\text{max}$ (W)           &     1           &   $x_d^\text{min}$ (m)     & -400           &$T$  (minute)           & $48$          \\ \hline
$P_{u,d,m}^n(r)$ (W)         &   0.1           &   $x_d^\text{max}$ (m)     &  400           & $T_n$ (minute)         & $8$          \\ \hline
$R^\text{th}$ (bps/Hz)       &   0.5           &   $y_d^\text{min}$ (m)     & -400           & $T_\text{move}$  (sec) & $30$          \\ \hline
$R^\text{BH}$ (bps/Hz)       &  10             &   $y_d^\text{max}$ (m)     &  400           & $B_0$  (kJ)            & $200$         \\ \hline
$N$                          &  6              &   $h$          ~~~~(m)     &  100           & $B_{00} $  (kJ)        & $400$         \\ \hline
$\rho_{o}$                   & 0.01            &   $v^\text{max}$ (m/s)     &   20           & $B_\text{charge}$ (kJ) & 25\%$B_0$             \\ \hline
$Q$                          & $10^6$          &   $B^\text{th}$ (kJ)       & 100            & $B_\text{PD}^\text{th}$(kJ)& 100         \\ \hline
\end{tabular}
\end{table}

\begin{figure}
            \centering
        \includegraphics[width=3.5in, height=1.8in]{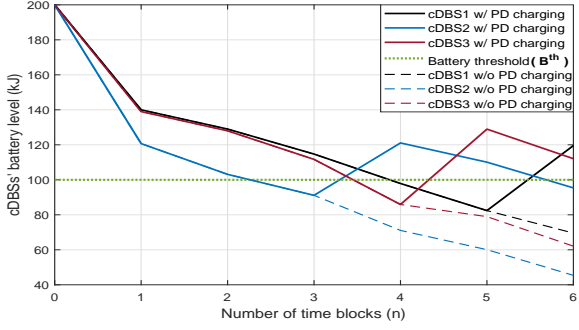}
        \caption{\,: cDBSs' battery levels with and without the PD.}
        \label{cDBSbattery}
\vspace*{-0.5cm}
\end{figure}

Fig. \ref{cDBSbattery} shows the battery level of each cDBS for time blocks from 0 to 6 where time block 0 is considered to be the initial state where all drones fly to reach the disaster area. Given that all cDBSs are initialized with a battery capacity of 200 kJ, the cDBSs are consuming their battery in hovering ,$E_\text{hov}$, moving, $E_\text{har}$, and in downlink transmission. From time block 0 to 1, all cDBSs are consuming high energy since they are crossing long distance to reach the disaster area. The solid lines represent the scenario where the PD is used. As it can be observed from the figure, all cDBSs are charged whenever their batteries' level is lower than $B^\text{th}$. At time block 4, cDBSs 1 and 3 curves are lower than $B^\text{th}$ although the lower battery were charged, the PD is choosing it randomly. If PD is not used, dashed curves, the cDBSs' grid will not be able to serve the disaster affected users more than 48 minutes.

It can be inferred from Fig. \ref{cDBSbattery} that the PD was not used until time block 3 and most of the cDBSs' battery level went near to the threshold level after time block 3, this motivates us to consider using an adaptive threshold level, $B^\text{th}$, which decreases as the time increases. This modification will be considered in the extended version of this paper.

\begin{figure}
            \centering
        \includegraphics[width=3.5in, height=1.8in]{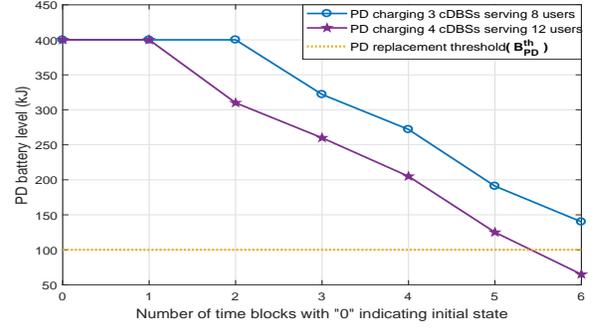}
        \caption{\,: PD's battery level for 3 cDBSs (8 users) and 4 cDBSs (12 users).}
        \label{PDbattery}
\vspace*{-0.5cm}
\end{figure}

Fig. \ref{PDbattery} shows the battery level of the PD versus the number of time blocks for 3 cDBSs/8 users and 4 cDBSs/12 users. For the PD serving 3 cDBSs which is related to the results in Fig. \ref{cDBSbattery}, the PD left its docking station with full battery towards cDBS 2 to charge it. During each time block the PD is charging the targeted cDBS with 50 kJ. For the 4 cDBS scenario, the battery level of the PD crossed the threshold level $B_\text{PD}^\text{th}$ in this case and based on our model, this PD will be replaced with a fully charged PD to take over the charging process and the depleted PD will return back to the docking station. This process will allow unlimited fly time for the flying cellular infrastructure. Note that if we provided 4 cDBSs to the scenario which is having 8 users, only three cDBSs will be used.

\section{Conclusion}
In this paper, we proposed a novel post-disaster rehabilitation framework for 4G/5G networks assisted by three different types of drones: 1) tethered Backhaul Drone (tBD) 2) untethered Powering Drone (PD) 3) untethered communication Drone Base-station (cDBS). This framework provides an unconstraint flying cellular infrastructure to any disaster area. An optimization problem is formulated where its objective is to minimize the consumed energy of the cDBSs. The optimization problem guarantees a minimum rate for each user in addition to finding the sub-optimal placement of the cDBSs and the time block to charge the cDBSs using PD. Results show that the minimum number of cDBSs is used. Also, the cDBSs are able to serve the users continually without the need to leave their location to charge their batteries due to the presence of the PD which is capable of charging cDBSs on the fly. 


\end{document}